\documentclass[a4paper,11pt]{article}
\usepackage{pos}
\pdfoutput=1
\usepackage{graphicx}
\usepackage{subfig}
\usepackage{color}
\usepackage{floatflt}

\usepackage{wrapfig}   
\usepackage{lipsum} 

\usepackage{multirow}
\usepackage{slashed}
\usepackage{verbatim} 
\usepackage{enumitem}
\usepackage{rotating}

\usepackage{multirow}
\usepackage{bm}
\usepackage[normalem]{ulem}
\usepackage{booktabs} 
\usepackage{tikz-cd}

\usepackage[fleqn,tbtags]{mathtools}

\usepackage{ytableau}

\usepackage{tikz}
\usepackage{diagbox}
\usetikzlibrary{positioning}
\usetikzlibrary{calc}
\usetikzlibrary{decorations.pathreplacing,calligraphy}

\usetikzlibrary{arrows}

\usepackage{xstring}
\usetikzlibrary{decorations.pathmorphing} 
\usetikzlibrary{decorations.markings} 
\usetikzlibrary{arrows} 
\usetikzlibrary{shapes} 
\usetikzlibrary{matrix} 
\usetikzlibrary{positioning} 
\usepackage[english]{babel} 
\usepackage[autostyle]{csquotes}

\newcommand{\be}{\begin{equation}}
\newcommand{\ee}{\end{equation}}
\newcommand{\ba}{\begin{aligned}}
\newcommand{\ea}{\end{aligned}}
\newcommand{\AdS}{\text{AdS}}
\newcommand{\Z}{\mathbb{Z}}
\newcommand{\C}{\mathbb{C}}

\title{Formal Theory at ICHEP 2024}

\author{Sakura Sch\"afer-Nameki}

\affiliation{Mathematical Institute, University of Oxford\\
Woodstock Road, OX2 6GG, Oxford, United Kingdom}

\abstract{These proceedings discuss some of the highlights of recent research in Formal Theory. The topics covered range from recent progress in scattering amplitudes, quantum gravity constraints on effective field theories, AdS/CFT, flat space holography, to generalized symmetries.}

\FullConference{42nd International Conference on High Energy Physics (ICHEP2024)\\
18-24 July 2024\\
Prague, Czech Republic\\}

\begin{document}
\maketitle


\section{What is "Formal Theory"?}

I was tasked with summarizing recent developments in "Formal Theory" for ICHEP 2024 in Prague. When I sought clarification on what precisely was meant by "Formal Theory," the response was somewhat ambiguous. It seemed that it would vaguely be anything theoretical that does not fit into the other well-defined categories of theoretical high-energy physics at ICHEP, or, more pointedly, "that part of theory which isn’t immediately useful for practical applications". So to start with lets clarify, what we will mean with 
"Formal Theory": the study that advances our fundamental understanding of Quantum Field Theory (QFT) and Quantum Gravity (QG), integrating insights from various areas of theoretical physics and mathematics. Naturally, any such theory should be seen with an eye toward potential phenomenological relevance. 

In view of the limited time and space allocated to Formal Theory at ICHEP\footnote{25 mins or 10 pages to cover hep-th, which in 2023 totals 3834 articles + 3452 cross-lists}, I have drawn inspiration 
 from some of the luminaries in the field, who have faced a similar predicament in the past \cite{Polchinski:2008ux, Quevedo:2022mji}, 
 and made an (evidently biased) selection: focusing on  topics that have seen significant recent progress (such as scattering amplitudes, precision holography, and quantum information aspects of QG) as well as rapidly growing fields that have garnered attention in recent years (such as flat-space holography and generalized symmetries).

\section{Scattering Amplitudes: QFT, String, Gravity}

The first formal theory topic of relevance to an ICHEP audience is that of scattering amplitudes. The progress is vast, and has been reviewed in the excellent 
 SageX review series \cite{Travaglini:2022uwo} and the Snowmass contributions \cite{Buonanno:2022pgc, Bern:2022jnl}. 

Indeed, one of the key theoretical challenges that the LHC poses is that of  precision prediction for QCD and electroweak sector scattering processes, at higher loop and higher external number of legs. This is pressing in light of the current limitations on QCD computations: 
4pt@3loop,  5pt@2loop (massless),  5pt@2loop 1-mass complete 1-mass pentagon functions \cite{Agarwal:2023suw, Caola:2021izf, Badger:2024sqv, Abreu:2023rco}.

Although much of these computations will require simply perseverance and computational power, there are useful insights from the structure of scattering amplitudes that are observed in idealized settings. One such framework is the supersymmetric cousin of QCD, 4d $\mathcal{N}=4$ supersymmetric Yang-Mills (SYM). Techniques to study amplitudes systematically using powerful mathematical ideas have emerged from the amplituhedron \cite{Arkani-Hamed:2013jha}, and recursion relations. The connection to string scattering amplitudes is sharpened by using in particular the holographic dual description of $\mathcal{N}=4$ SYM as string theory on $\text{AdS}_5\times S^5$. 

Another very quickly evolving field is the relation between Yang-Mills scattering and scattering in gravity -- the main tool being the double-copy. Using color-kinematics duality allows relating scattering in gauge to gravity theories. 

As most of the {\bf Formal Theory parallel sessions at ICHEP 2024} were on the topic of scattering amplitudes, and will cover all the above range of approaches and recent results, we will leave a thorough discussion of scattering amplitudes to these contributions.

\section{Quantum Gravity Constraints on IR Physics}

Quantum Gravity (QG) constraints on low energy effective field theories (EFTs) are a major area of exploration within string theory, called the swampland program. 
This has been a very active field, reviewed multiply in the following works \cite{Brennan:2017rbf, Palti:2019pca, vanBeest:2021lhn, Harlow:2022ich, Reece:2023czb}. 
Here, we will highlight a few recent results that are particularly noteworthy, as they demonstrate the increasing precision of statements within the field, and the renewed interest in non-supersymmetric theories. 

The traditional approach to string phenomenology, known as top-down, starts with a 
10d string theory $\mathcal{T}_{10}$ compactified to 4d, where it gives rise at low energies to an EFT, e.g. SM spectrum $+ X$ (where $X$ is usually $N=1$ susy, exotics). Schematically this takes the form
$\mathbb{R}^{1,9} \  \to \  \mathbb{R}^{1,3} \times  M_6$, 
where $M_6$ is a  compact space, whose structure crucially determines the properties of the EFT.
\begin{wrapfigure}{l}{0.45\textwidth}
    \centering
   \includegraphics*[width= 4.5cm]{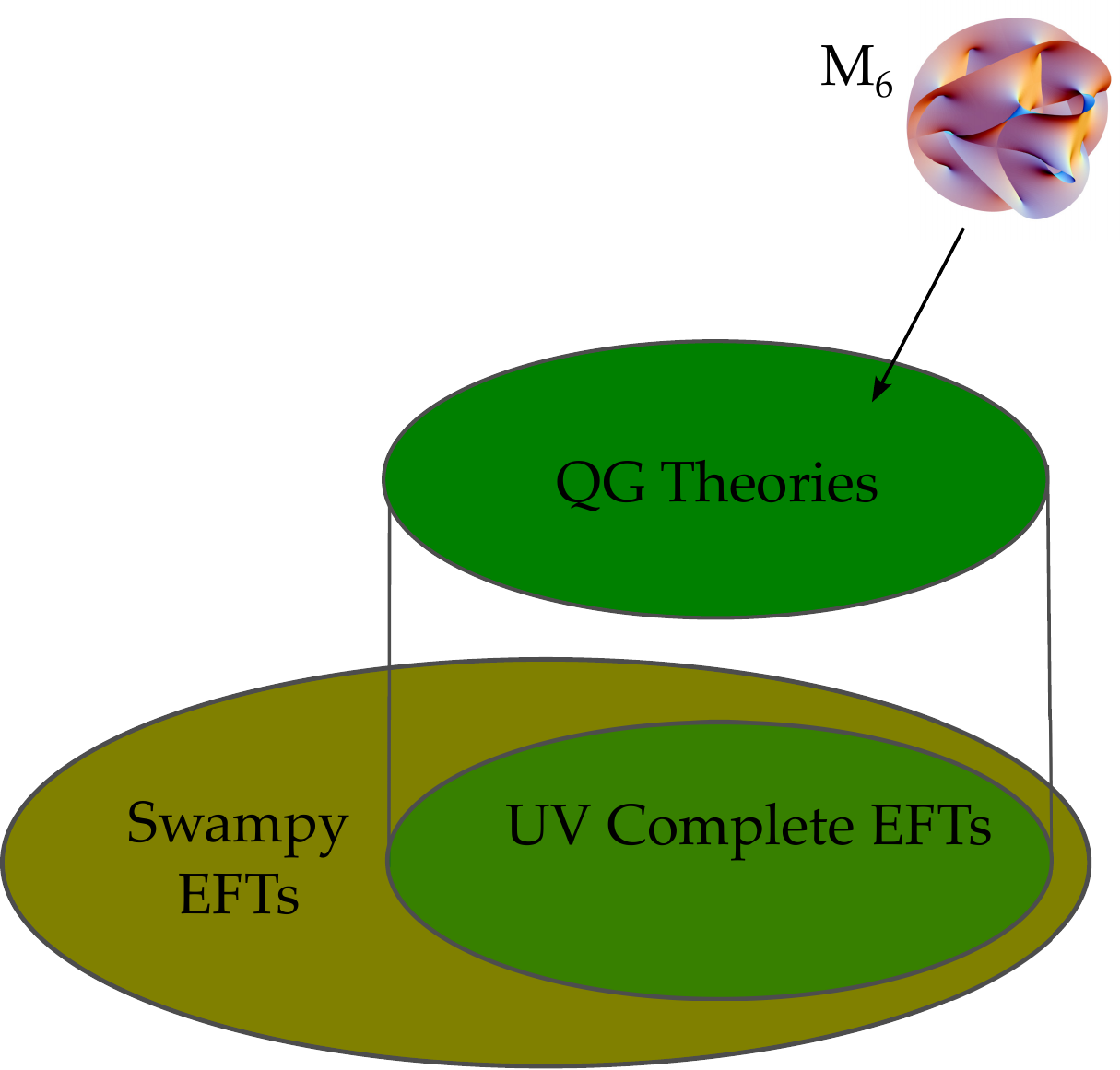}
    \caption{Swampland and Landscape. \label{fig:Swamp}}
\end{wrapfigure}
This defines at low energies an {EFT ($\mathcal{T}_{10}, M_6$)}, which by construction has a consistent UV completion. An alternative, bottom up, perspective is guiding the swampland program. 
The key questions driving this program are:
\begin{itemize}
    \item Can a given EFT be embedded into a consistent theory of quantum gravity? If not it is in the swampland, see figure \ref{fig:Swamp}.
    \item Is there a universality to string theory, such that all consistent quantum gravity theories are derived from string theory?
\end{itemize}
The swampland conjectures propose criteria such that EFTs must satisfy to be consistent with quantum gravity. Three significant conjectures are: 
\begin{itemize}
    \item \textbf{Distance Conjecture}: As one moves towards infinite distance in field space, an infinite tower of light states emerges, causing the breakdown of the EFT.
    \item \textbf{Weak Gravity Conjecture}: Gravity must be the weakest force, implying that there must be a particle with a charge-to-mass ratio greater than that of a black hole.
    \item \textbf{No global Symmetry Conjecture}: all global symmetries in QG have to be gauged or broken.
\end{itemize}
The key distinction in the progress of recent years is the ability to make precise statements. E.g.
it is proven that there is string universality (i.e. all consistent EFTs are UV completed in string theory) for 
 8d and 9d abelian gauge theories with 16 supercharges \cite{Montero:2020icj}. Recent advances continue to push these precision statements to lower dimensions and lower  supersymmetry. 

Some interesting progress has been made also in terms of non-supersymmetric string theories. As is known -- since a long time ago -- there are three non-supersymmetric, tachyon-free
10d string theories \cite{Dixon:1986iz, Sugimoto:1999tx, Sagnotti:1995ga}. Their properties are summarized in table \ref{tab:nonsusy}.
Recently these models have been revisited and various statements regarding their uniqueness and consistency have been made:  
\begin{enumerate}
\item Completeness: \cite{BoyleSmith:2023xkd} these are all the non-supersymmetric 10d heterotic string theories.
\item \cite{Basile:2023knk} showed using cobordism theory,  that all local and global anomalies cancel.
\item By compactification, there are  new $d<10$ non-supersymmetric theories \cite{Gkountoumis:2023fym, Basaad:2024lno, Baykara:2024tjr}.
\end{enumerate}

\begin{table}
\centering
\begin{tabular}{c|c|c}\hline
$SO(16)^2$  Het  (1987) \cite{Dixon:1986iz} & $Sp(16)$ Sugimoto (1999) \cite{Sugimoto:1999tx} & 0'B Sagnotti  (1995) \cite{Sagnotti:1995ga}  \cr \hline\hline
  $E_8\times E_8$ Het/$(-1)^F$& IIB + $O9^+$ + 32 ${\overline{D9}}$    &  IIB/$(-1)^F/\Omega$ \cr 
Heterotic $g, \phi, B$  & Closed strings: $\mathcal{N}=1$ susy &Closed: metric dilaton   \cr 
No gravitino  & Open strings: non-susy & non-susy  \cr 
$SO(16)^2$ gauge group& \qquad $Sp(16)$ gauge group&  $U(32)$ gauge group\cr \hline
\end{tabular}
\caption{Three 10d non-supersymmetric, tachyon-free string theories. \label{tab:nonsusy}}
\end{table}
Clearly much more needs to be understood in terms of stability of non-supersymmetric vacua, improving their phenomenological appeal (e.g.  gauge groups, matter content, interactions). However the main, noteworthy recent progress is that for certain classes of EFTs, our theoretical understanding has evolved to make very precise, universal statements.

\section{Holography}

Holography continues to be one of the most vibrant and active fields in formal theory. It is the classic example of a gauge/gravity or strong/weak duality and provides a conceptual and computational window into quantum gravity and strongly-coupled QFTs alike.
There are three developments that I will highlight here, where progress in the past few years has been outstanding. 

\subsection{AdS/CFT Precision Holography}

Holography provides a profound connection between quantum gravity and quantum field theory, particularly through the AdS/CFT correspondence, which posits an exact duality between a QG in $d+1$ dimensional AdS space and a conformal field theory (CFT) on its $d$-dimensional boundary. 
\begin{wrapfigure}{l}{0.5\textwidth}
    \centering
\includegraphics*[width=7cm]{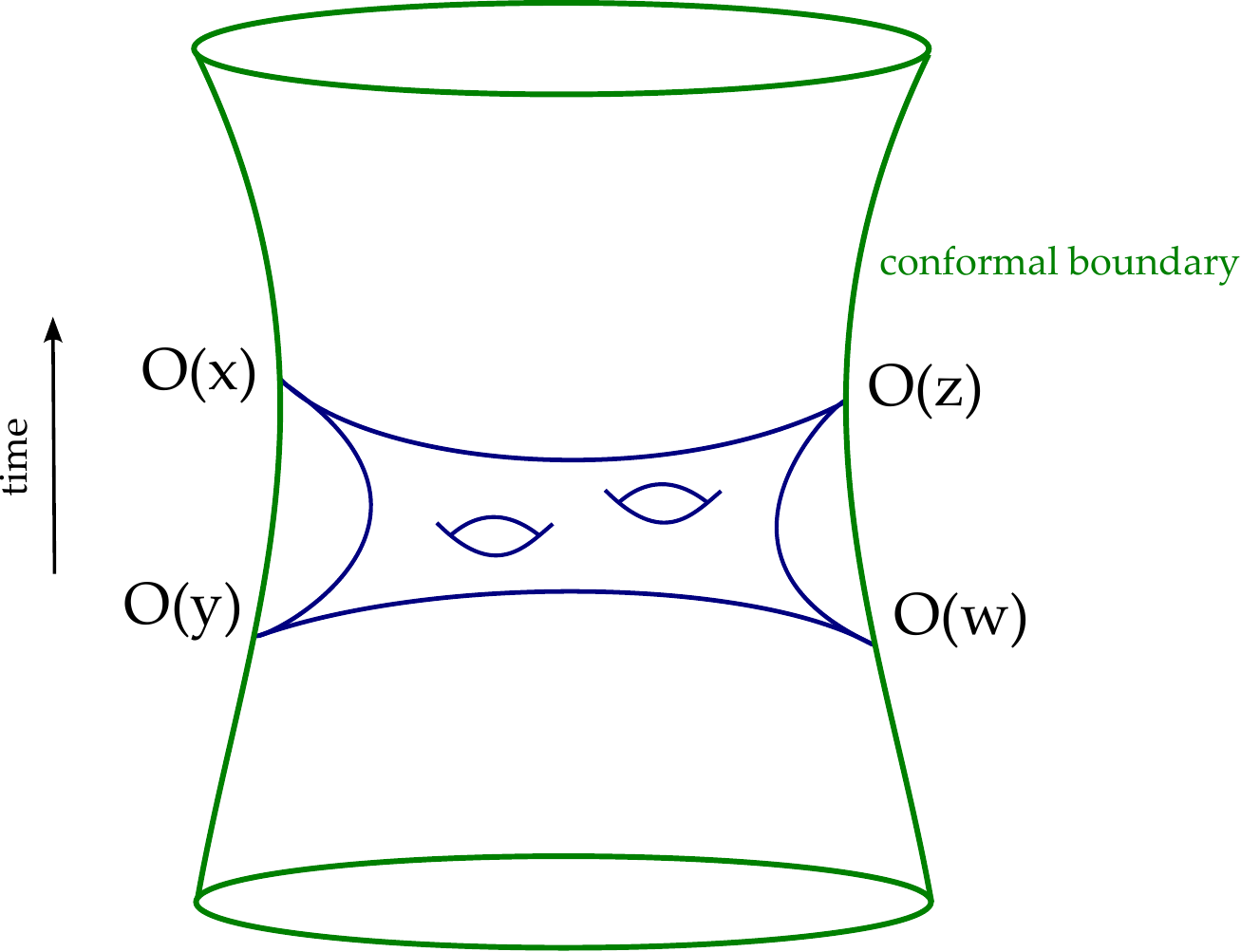}
    \caption{Scattering in AdS and correlators in the boundary CFT.\label{fig:bounty}}
\end{wrapfigure}
The classic example is that of type IIB string theory on $\text{AdS}_5\times S^5$ and 4d $\mathcal{N}=4$ $SU(N_c)$ SYM.
This sets up a precision (in coupling constants on both sides) lab for quantum gravity and strongly-coupled CFTs alike. The map is controlled by a well-founded dictionary: e.g.  
the conformal symmetry is identified with the isometries of AdS-space. 
Key perturbation parameters are identified: e.g. the  string coupling $g_s = {1\over N_c}$ and string length scale ($R$ radius of the sphere/AdS)  ${R^2 \over \alpha'} = \sqrt{\lambda} = \sqrt{g_{YM}^2 N_c}$, where $\lambda$ is the 't Hooft coupling of the CFT. 

Let us highlight two notable developments: 
First of all the $\text{AdS}_3/\text{CFT}_2$ is now understood to be an exact duality. More precisely for the $\text{AdS}_3\times S^3\times T^4$ background of Type IIB string theory with 1 unit of NSNS flux, an exact correspondence to the  2d CFT 
 given by the symmetric orbifold of the torus $T^4$ CFT, $\text{Sym}^N (T^4)$, was shown \cite{Eberhardt:2019ywk}.

Secondly, one of the classic string theory computations of string scattering in flat space has been extended to $\AdS_5\times S^5$ (see figure \ref{fig:bounty}) -- which is a notoriously difficult string theory background. 
The philosophy here is to use insights and consistency conditions from QFT to reconstruct string amplitudes, such as the  4-graviton scattering in $\AdS_5\times S^5$:
\be
\includegraphics*[width = 12cm]{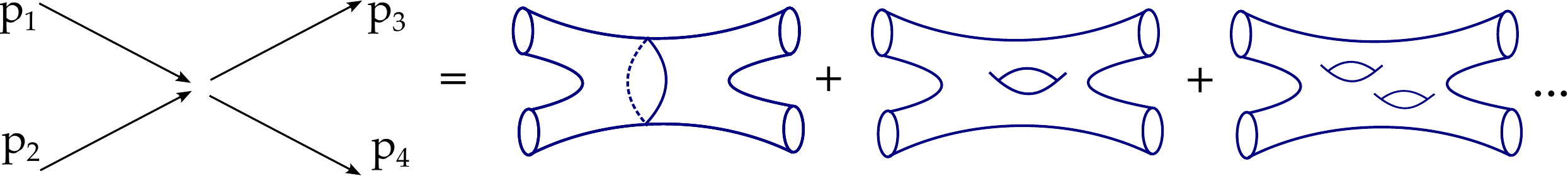} 
\ee
which has an expansion in {string-loops, i.e. $1/N_c$}. Even the  tree-level amplitude in curved spacetimes (with fluxes) such as $\AdS_5\times S^5$ in notoriously hard. 
This so-called Virasoro-Shapiro amplitude has an expansion in {$\alpha'$ or 't Hooft coupling $1/\sqrt{\lambda}$} 
\be
A(S, T)=A^{(0)}(S, T)+{ A^{(1)}(S, T) \over \sqrt{\lambda}}+\cdots  \,,
\ee
where the flat space amplitude is the well-known expression
\be
A^{(0)}(S, T)= \frac{1}{U^2} \int d^2 z|z|^{-2 S-2}|1-z|^{-2 T-2} = -\frac{\Gamma(-S) \Gamma(-T) \Gamma(-U)}{\Gamma(S+1) \Gamma(T+1) \Gamma(U+1)} \,.
\ee
Using insights from  conformal bootstrap and supersymmetric  localization,  integrability 
and number-theory, \cite{Alday:2023mvu}
conjectured the subleading terms and wrote them in terms of a world-sheet-type integral with insertions:
\be
{A_4^{A d S}(S, T) \sim \int d^2 z|z|^{-2 S}|1-z|^{-2 T} W_0(z, \bar{z})\left(1+\frac{S^2}{R^2} W_3(z, \bar{z})+\frac{S^4}{R^4} W_6(z, \bar{z})+\cdots\right)}
\,, \ee
where {$W_n$'s are single-valued polylogarithms of weight $n$}.
Recent progress on  String Field Theory in backgrounds like $\AdS_5\times S^5$ should soon be able to test this. 

In summary, the AdS/CFT correspondence is now a precision lab for strongly-coupled QFTs and quantum gravity. Although these are again not  real world holographic duals,  many important lessons can learned from these computations (see also the later section on flat space holography).

\subsection{Quantum Information and Quantum Gravity}

One of the key lessons of the past few years has been that there is a very close connection between QG and quantum information theory. 
For instance, a key observation is that  ordinary quantum systems, e.g. spin chains, can have emergent gravitational properties. This was profoundly investigated and shown 
for the  Sachdev-Ye-Kitaev (SYK) model of $N$ Majorana fermions in 1+1d with random couplings
\be
  H_{\text{SYK}} =  \sum_{i_1, i_2, \ldots, i_q} J_{i_1 i_2 \cdots i_q} \psi_{i_1} \psi_{i_2} \cdots \psi_{i_q}  \,.
\ee
At large $N$, this has emergent gravity (Jackiew-Teitelboim (JT) gravity) \cite{kitaev2015talks, Maldacena:2016upp, Jensen:2016pah, Engelsoy:2016xyb}. The SYK model has  emergent conformal symmetry with central charge $ c \approx \frac{N^2}{2}$. Furthermore, the entropy is akin to that of a black holes: $S \approx \frac{N^2}{4} \log T$. 

Quantum information concepts have become more generally key tools in hep-th: 
starting with spacetime emergence from quantum entanglement ( ER= EPR), where entangled particles are connected by wormholes.
In holography, the entanglement entropy is given by an area (minimal surface)$/(4 G_N)$. 
And most relevant for the ICHEP audience, this has even applications in hep-ph/ex, with the measurement of top-quark entanglement \cite{ATLAS:2023fsd}, as reviewed in {\bf Dunford’s ICHEP plenary talk on ATLAS}.

\subsection{Flat-Space Holography}

A natural question is whether there is a holographic principle for asymptotically flat spacetimes (AFS), which has potential applications (via scattering problems in AFS) to collider physics and astro-physical settings. Indeed two recent proposals have been put forward: 
\begin{itemize}
    \item \textbf{Celestial Holography}: 
    $(d+{2})$-dim AFS dual to a $d$-dim  CFT on the celestial sphere at future/past null-infinity $\mathscr{I}^\pm$. In particular 4d gravity is dual to 2d CFT on a celestial sphere $S^2$. For recent overviews of this rapidly developing field see \cite{Raclariu:2021zjz, Pasterski:2021raf,Pasterski:2021rjz}.
    \item \textbf{Carrollian Holography}:
        A duality between $(d+1)$-dimensional gravity and $d$-dimensional Carrollian field theories. 4d gravity is dual to a 3d Carrollian\footnote{This limit has a peculiar causal structure, like in Lewis {Carroll}'s "Through the Looking Glass"} (sending the speed of light $c\to 0$) field theory at null infinity  \cite{Barnich:2006av, Donnay:2022aba}.
        
\end{itemize}

\subsubsection{Celestial Holography}

\begin{wrapfigure}{l}{0.4\textwidth}
    \centering
\includegraphics*[width = 3.5cm]{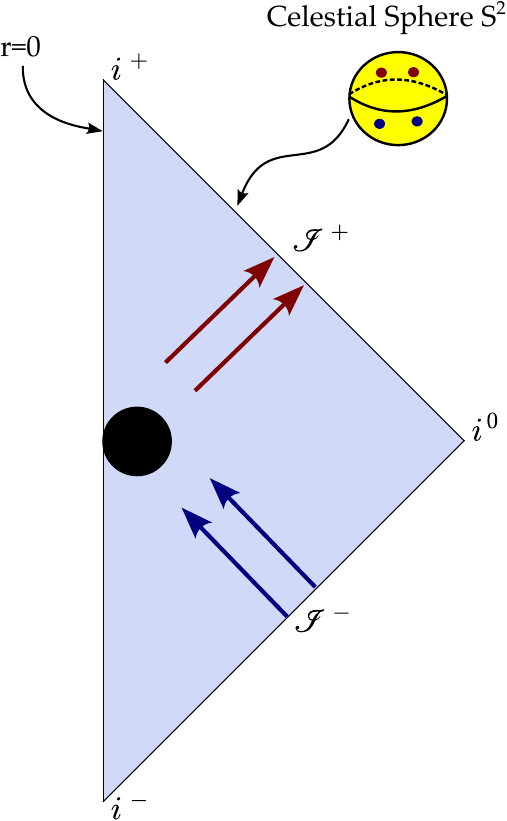} 
\caption{Penrose diagram for scattering in asymptotic flat space as a correlator in the celestial CFT.\\ }
\end{wrapfigure}
As with any holography, the first point is to identify the symmetries: the 4d Lorentz group is identified with the conformal group on $S^2 \subset$ null infinity $\mathscr{I} \cong \mathbb{R}\times S^2$.
Most crucially, scattering amplitudes in 4d become correlators in the 2d celestial CFT (CCFT):
the S-matrix between $\mathscr{I}^- $  and $\mathscr{I}^+$ becomes a  correlator in the  2d CCFT:
$$ 
\left\langle \textcolor{red}{p_n^{\text {out }}} \ldots|\mathcal{S}| \ldots \textcolor{blue}{p_1^{\text {in }}}\right\rangle 
\longleftrightarrow \quad\left\langle\mathcal{O}_{\Delta_1}^{-}\left(x_1\right) \ldots \mathcal{O}_{\Delta_n}^{+}\left(x_n\right)\right\rangle
$$
Furthermore, {gravitational memory effects} (permanent displacement of test particles (detectors) due to the passage of gravitational waves), correspond to change of state in CCFT \cite{Strominger:2014pwa}. 
 {The Bondi-Metzner-Sachs (BMS)  symmetry group},  which describes the asymptotic symmetries at null infinity,  including supertranslations and superrotations, implies via Ward identities, soft graviton theorems in the bulk. Futhermore, it is 
enhanced to the so-called $w_{1+\infty}$ algebra  (by including an infinite tower of higher-spin fields), potentially offering new observables in flat space.

\subsubsection{Carrollian Holography}

\begin{figure}[h]
\centering
\includegraphics*[width = 4cm]{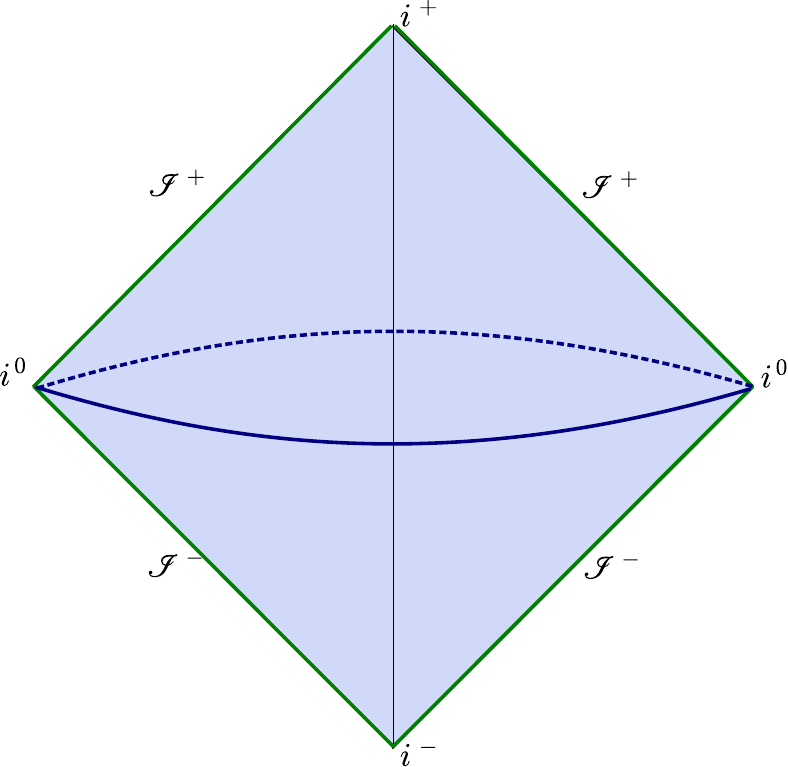}\qquad \qquad  
\includegraphics*[width = 3.5cm]{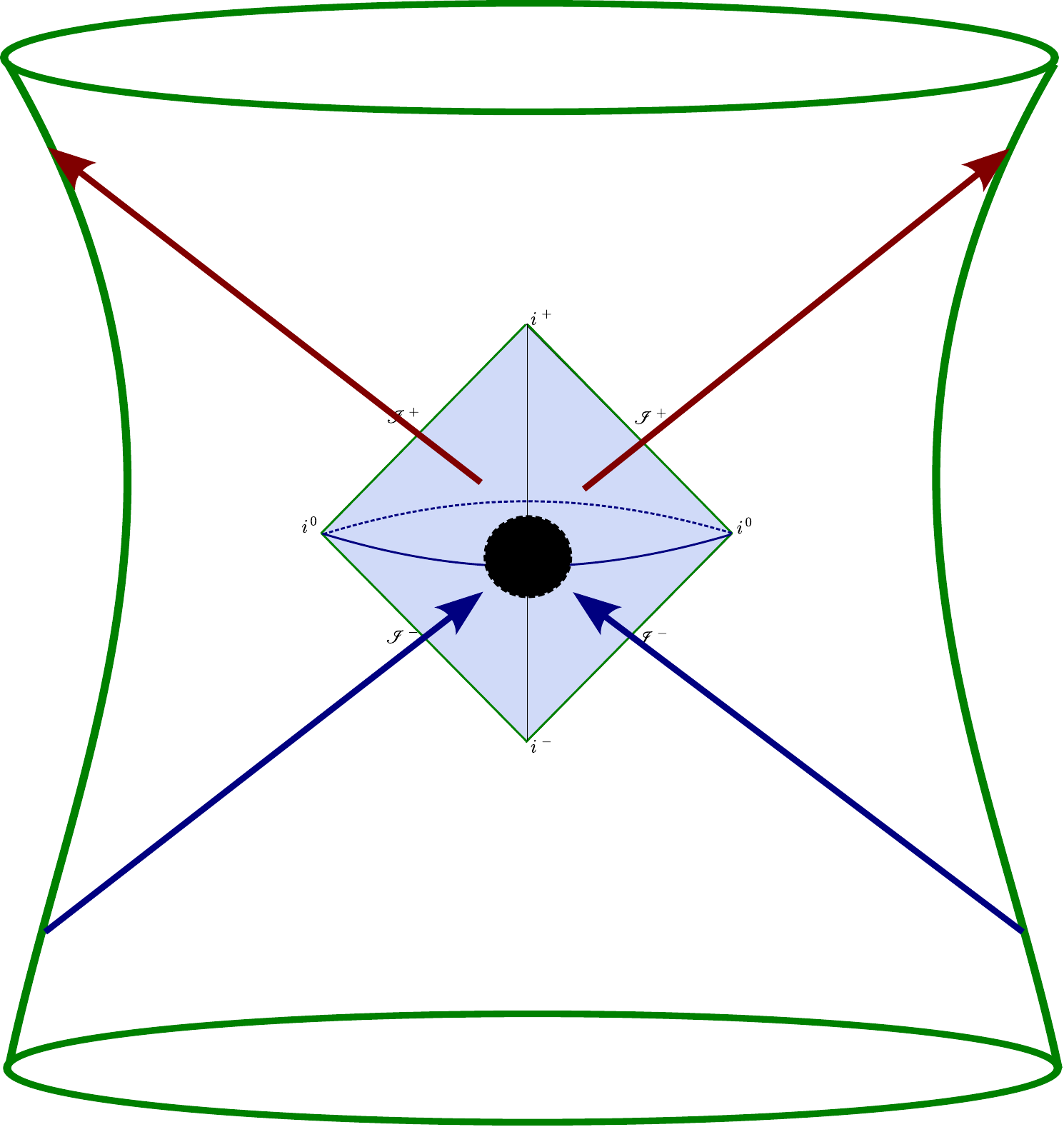} 
\caption{Penrose diagram for Carollian Holography (left). Carrollian limit of AdS/CFT. \label{fig:bla}}
\end{figure}

In this case the proposed correspondence is between a $d+1$ dimensional AFS gravity dual to $d$-dimensional  Carrollian field theory living on ${\mathscr{I} \simeq \mathbb{R} \times S^{d-1}}$. 
As there are no reviews on this topic, a few foundational papers should be mentioned: Carrollian holography has been discussed for 3d gravity \cite{Barnich:2006av, Bagchi:2010zz, Barnich:2012xq, Bagchi:2012xr} and 4d in \cite{Arcioni:2003xx, Barnich:2010eb, Duval:2014uva, Ciambelli:2018wre, Donnay:2022aba}. 
The relation to scattering amplitudes in 4d and Carrollian holography is  explored \cite{Donnay:2022wvx, Mason:2023mti}. 
Some of the basic features of this flat space holography are as follows: 
The symmetry of the theory at null infinity is obtained in a Carrollian limit $c\to 0$ of the conformal group, where the  lightcones collapse along  the $t$-axis. Spacetime events are therefore causally disconnected.
The BMS algebra maps to the conformal Carrollian algebra. 

 As the codimension of the boundary is one, just as in AdS/CFT, one can take the $c\to 0$ of relativistic CFTs dual to a local flat-space patch in AdS (figure \ref{fig:bla}). Using AdS/CFT holographic correlators, then allows one to derive Carrollian amplitudes \cite{Alday:2024yyj}. 

In summary, this is a very active area of research, with huge potential. There are many open questions, that however need to be addressed before AFS holography becomes of similar standing as AdS/CFT. E.g. even the precise definition of the CCFT remains elusive and requires substantially more work in order to pin it down precisely.

\section{Generalized Symmetries}

The final topic\footnote{, which is the actual research topic of the author,} is that of generalized/non-invertible/categorical symmetries. 
For  reviews on this rapidly evolving field at the intersection of high energy theory, condensed matter theory and mathematics, see \cite{Schafer-Nameki:2023jdn, Brennan:2023mmt, Bhardwaj:2023kri,  Shao:2023gho, Luo:2023ive}. 

Symmetries are undoubtedly a cornerstone of any theory, formal or not, high or low energy. They e.g.  constrain the spectrum of particles, interactions, provide constraints on the IR through anomalies. 
Usually, symmetries are understood to be groups, i.e. there is a composition $\cdot$ (multiplication) and for each symmetry generator there is an inverse. Also, ordinarily symmetries will act on point-like objects (local operators), which transform in (unitary) representations of the group.

Starting with the seminal work \cite{Gaiotto:2014kfa} the past 10 years revolutionized our understanding of what a {\bf global symmetry} is. The key insight underlying this development is as follows: 
{\bf a global symmetry is simply a topological\footnote{Topological meaning independent of precise metric shape, only dependent on the topology.} operator}.  There are two stages of generalization:
\begin{itemize}
\item {\bf Higher-form symmetries:} \cite{Gaiotto:2014kfa} Symmetries that act on extended operators (a $p$-dimensional operator is charged under a  $p$-form symmetry). 
The main example is that of 
line operators charged under a 1-form symmetry (which is often also referred to as the center symmetry, e.g. $\mathbb{Z}_N$ for $SU(N)$ gauge groups). The charge is computed by linking the 1-form symmetry generator, in 4d a topological surface operator, with the line: 
\be
{\begin{tikzpicture}
\draw [thick](-4,0) -- (-2,0);
\node[right] at (-2,0) {Line operator $L_{1}$};
\draw [ultra thick,white](-3.4,0) -- (-3.2,0);
\draw [thick,blue] (-3,0) ellipse (0.3 and 0.7);
\draw [ultra thick,white](-2.7,-0.1) -- (-2.7,0.1);
\draw [dashed,thick](-2.7396,0) -- (-2.6396,0);
\node[blue] at (-3,-1) {2d 1-form symmetry generator};
\end{tikzpicture}
}
\ee
Physically 1-form symmetries are crucial in characterizing confinement: the confining phase is the symmetric phase, whereas the deconfined phase is the spontaneously broken symmetry (SSB) phase. The order parameter is the vev of the Wilson line operator that is charged under the 1-form symmetry. 

\item {\bf Non-invertible (or categorical) symmetries:} The constructions in \cite{Kaidi:2021xfk,Choi:2021kmx, Bhardwaj:2022yxj} showed that one can furthermore relax the existence of inverses. If
$a, b$ are symmetry generators of a non-invertible symmetry then their composition will generically be 
\be
{a \cdot b=  n_{c_1} c_1 + \cdots + n_{c_k} c_k }\,,\qquad c_i \in \mathcal{S}\,, 
\ee
and in the simplest instance $n_{c_i}$ are non-negative integers. More generally one can combine this with higher-form symmetries to get higher fusion category symmetries. 
\end{itemize}
Although this may all sound like fanciful mathematics, it turns out that non-invertible (categorical) symmetries are ubiquitous in QFT, including the Standard Model of Particle Physics
\cite{Choi:2022jqy, Cordova:2022ieu}!

\subsection{Non-Invertible Symmetries in the Ising CFT}

The transverse field  Ising model spin chain is the simplest setup where non-invertible symmetries can be discussed. It has a Hilbert space  $\mathcal{H}= (\C^2)^L$, where $L$ is the length of the chain, with nearest neighbor Hamiltonian
\be
H=-\sum_{j} \sigma^z_j \sigma^z_{j+1}- g \sum_j \sigma^x_j \,.
\ee
There is a $\Z_2$ spin flip symmetry  $\eta = \prod_j \sigma^x_j$. The phase space is schematically given by the following diagram: 
\be
\begin{tikzpicture}
{\node at (5,0.2) {$g$};
\draw [-stealth, thick](-1,0) -- (5,0);
\draw [black, thick, fill= black] (0, 0) ellipse (0.05 and  0.05);
\draw [blue, thick, fill=blue] (2, 0) ellipse (0.05 and  0.05);
\node[below] at (0,0) {$g=0$};
\node[below] at (2,0) {$g=1$};
\node[above] at (0.4,0.5) {$\text{ordered}$};
\node[above, rotate=90, blue] at (2.3,1) {$\text{Ising CFT}$};
\node[above] at (3.5,0.5) {$\text{disordered}$};
}
\end{tikzpicture}
\ee
There are two gapped phases: $g=0$ with two ground states (describing the $\Z_2$ SSB) and the $g \gg 1$ phase that is the $\Z_2$ symmetric phase. The critical Ising CFT is obtained by setting $g=1$. 
This spin-chain has a Kramers-Wannier duality, which maps  
$\sigma^x_i \to \sigma^z_j \sigma^z_{j+1}$ and $\sigma^z_j \sigma^z_{j+1} \to \sigma^x_{j+1}$,  or equivalently   $g\to g^{-1}$. 
At $g=1$, which is the critical Ising CFT, this becomes a {\bf Kramers-Wannier duality symmetry $N$}, which satisfies the following relation 
\be
{N \cdot N = 1+ \eta}\,,
\ee
which is precisely the composition of a non-invertible symmetry.

\subsection{Non-Invertible Symmetries in $d=4$}

Following the construction before of the non-invertible symmetry  of the Ising CFT, \cite{Kaidi:2021xfk,Choi:2021kmx}
generalized this to 4d: when a theory has a duality symmetry $D$, it can -- for specific choices of parameters -- become a self-duality 0-form symmetry, with non-invertible composition. An example is the 4d $\mathcal{N}=1$ SYM theory with gauge group  $SO(3)$.

Another construction is that of gauging charge conjugation (cc) \cite{Bhardwaj:2022yxj}, e.g. the  $O(2)=U(1)/\Z_2^{\text{cc}}$ gauge theory has non-invertible symmetries.
There is a 1-form symmetry, generated by $D_\alpha:= e^{i \alpha \int \star F} $, with the field strength $F=dA$, which is acted upon by $\Z_2^{\text{cc}}:\  D_\alpha  \to  D_{-\alpha} $:
\be
\begin{tikzpicture}
 \scalebox{0.9}{
\begin{scope}[shift={(0,-4)}]
\node at (-3, 3) {${D_\alpha^{\text{inv}}= D_\alpha \oplus D_{-\alpha}}$};
\draw [fill=cyan,opacity=0.3] (0,0) -- (2,0) -- (2,3.5) -- (0,3.5) -- (0,0);
\draw [black, thick] (0,0) -- (2,0) -- (2,3.5) -- (0,3.5) -- (0,0);
\node [black, thick] at (1,2.5) {$\Z_2^{cc}$};
\draw [thick,white](0,1.1) -- (0,0.9);
\node [red, thick] at (-1.5,0) {$D_\alpha$};
\node [red, thick] at (3.5,3) {$D_{-\alpha}$};
\draw [red, thick] (-1,0.5) -- (1,1.5);
\draw [red,fill=red] (1,1.5) ellipse (0.05 and 0.05);
\draw [red, thick, dashed] (1,1.5) -- (1.95,2);
\begin{scope}[shift={(0,0.05)}]
\draw [red, thick] (2.05,2) -- (3,2.5);
\end{scope}
\end{scope}}
\end{tikzpicture}
\ee
The invariant combination $D_\alpha^{\text{inv}}$ then has a non-invertible fusion $D_\alpha^{\text{inv}} \cdot D_\alpha^{\text{inv}} = D_{0} + D_{2\alpha}$ for $\alpha\not= \pi/2$ (for a more in depth discussion of the fusions see \cite{Bhardwaj:2022yxj, Schafer-Nameki:2023jdn}).

\subsection{Non-Invertible Symmetries from ABJ Anomalies}

An Adler-Bell-Jackiw (ABJ) anomaly is the non-conservation of an axial $U(1)_A$ symmetry in the quantum theory. E.g. 
4d QED with massless charge $q=1$ Dirac fermion $\Psi$, with Lagrangian 
\be 
\mathcal{L}_{\text{QED} + \Psi} = \frac{1}{4 e^2} F_{\mu \nu} F^{\mu \nu}+i \overline{\Psi}\left(\partial_\mu-i A_\mu\right) \gamma^\mu \Psi\,,
\ee
has an axial current
$j_\mu=\frac{1}{2} \bar{\Psi} \gamma_5 \gamma_\mu \Psi$, which is not conserved
$
{d \star j=\frac{1}{8 \pi^2} F \wedge F } \not=0$. 
Instead of interpreting this as the absence of the axial symmetry in the quantum theory, one can instead consider defining a symmetry operator as follows: consider a rotation angle $1/N$ then naively we would write the symmetry generator as $\int[D a] \exp \left(\int_{M_3} \frac{2 \pi i}{N} \star j\right)$, where $a$ is a dynamical gauge field. This is not topoglocal due to the ABJ anomaly. Instead consider 
\be 
\mathcal{N}_{\frac{1}{N}}(M_3)=\int[D a] \exp \left(\int_{M_3} \frac{2 \pi i}{N} \star j + {\frac{i N}{4 \pi} a d a+\frac{i}{2 \pi} a d A}\right) \,
\ee
where we added a coupling to a 3d topological field theory (TQFT), which ensures that  $\mathcal{N}$ is topological, and thus generates a symmetry. 
However this comes at the price of non-invertibility:
\be \mathcal{N}_{\frac{1}{N}} \times {\mathcal{N}^\dagger_{\frac{1}{N}}}  =\mathcal{C} =\text{condensation defect for 1-form symmetry}\not= {\bf 1} \,,
\ee
where the right hand side is a 3d topological defect, on which the 1-form symmetry is gauged \cite{Roumpedakis:2022aik, Bhardwaj:2022lsg}. 
This construction of non-invertible symmetries from ABJ anomalies has many applications: starting with constraints on pion decays \cite{Choi:2022jqy, Cordova:2022ieu}, and some models of neutrino masses \cite{Cordova:2022fhg} (where there is  Z' that has a non-invertible chiral symmetry).

\subsection{Physical Implications of Non-Invertible Symmetries}

How are non-invertible symmetries distinguished from group ones? The first hint is in the representation theory: they generically map genuine (untwisted sector operators, not attached to any topological defect) to non-genuine (twisted sector, attached to topological defects) \cite{Frohlich:2009gb, Lin:2022dhv, Bhardwaj:2023ayw, Bartsch:2023pzl}: 
\be
\begin{tabular}{c c}
{\bf Example: 2d Ising CFT} $ N^2 = 1+ \eta $ 
& \qquad \qquad 
{\bf Example: Witten effect.} 4d $SO(3)$ SYM\cr 
\scalebox{0.65}{\begin{tikzpicture}
\begin{scope}[shift={(0,0)}]
\draw [blue, thick] (0,0)-- (0,4);
\draw [red ,fill= red ] (-1,2) ellipse (0.07 and 0.07);
\node[above, red ] at (-1,2) {$\sigma$};
\node[below, blue] at (0,0) {${N}$};
\end{scope}
 \begin{scope}[shift={(3,0)}]
 \draw[thick, ->] (-2, 2.2) -- (-1, 2.2);
 \draw[blue, thick] (0,0) -- (0,4);
  \draw[teal, thick] (0,2.2) -- (1.7,2.2);
 \draw [red ,fill= red ] (1.7, 2.2) ellipse (0.07 and 0.07);
\node[above,  red ] at (1.7, 2.2) {$\mu$};
\node[below, blue] at (0,0) {${N}$};
\node[above, teal] at (0.8,2.2) {\text{$\eta$}};
 \end{scope} 
\end{tikzpicture}}
& 
\begin{tikzpicture}
\scalebox{0.9}{
\begin{scope}[shift={(0,0)}]
\draw [red] (0,1) ellipse (0.1 and 0.5);
\draw [thick,blue, fill= blue, opacity= 0.2]
(1,0) -- (1,2) -- (1.5, 2.2) -- (1.5, 0.2) -- (1,0);
\node[red] at (0, -0.5) {$\mathbf{H}$};
\node[blue] at (1.2,-0.5) {$\mathcal{N}$};
\node[black] at (2.5, 1) {$\longrightarrow$} ;
\end{scope}
\begin{scope}[shift={(3.8,0)}]
\draw [thick, teal] (0.25, 1.5) -- (1.5, 1.5);
\draw [thick, teal] (0.25, 0.5) -- (1.5, 0.5);
\draw [fill= teal, opacity=1] 
(0.25, 1.5) -- (1.5, 1.5) -- (1.5, 0.5) -- (0.25, 0.5) -- (0.25, 1.5);
\draw [thick,blue, fill= blue, opacity= 0.2]
(0,0) -- (0,2) -- (0.5, 2.2) -- (0.5, 0.2) -- (0,0);
\draw [teal, fill= teal, opacity= 1] (0.25,1) ellipse (0.1 and 0.5);
\draw [-stealth,  fill = teal, opacity= 1] (1.5,1) ellipse (0.1 and 0.5);
\draw [thick, red] (1.5,1) ellipse (0.1 and 0.5);
\node[red] at (1.5, -0.5) {$\mathbf{H}$};
\node[blue] at (0.2,-0.5) {$\mathcal{N}$};
\end{scope}}
\end{tikzpicture}
\end{tabular}
\ee
On the left we show the passage of spin operator, through the non-invertible line, which maps to a twisted sector operator.
On the right we depict the action of the non-invertible symmetry of $SO(3)$ $\mathcal{N}=1$ SYM, which maps the 't Hooft loop to a flux attached 't Hooft loop. 

Non-invertible symmetries in 1+1d imply {\bf modified crossing relations for the S-matrix} \cite{Copetti:2024rqj}. {\bf Scattering electrically charged, massless fermions off
magnetic monopoles} (or 't Hooft line) has an explanation in terms of generalized symmetries \cite{vanBeest:2023dbu, vanBeest:2023mbs}, where a genuine particle scatters off, and gets emitted with a topological defect attachment. 

Finally, perhaps most importantly for condensed matter physics, this results in a new Landau paradigm for phases of matter. The standard paradigm states that second order phase transitions are symmetry breaking transitions -- where the symmetry is a group, and the phases are labeled by subgroups (and cocycles). A {\bf categorical Landau paradigm} was proposed in \cite{Bhardwaj:2023fca}, which provides a systematic way to classify gapped phases and gapless phase transitions using the symmetry topological field theory (SymTFT) \cite{Gaiotto:2020iye, Apruzzi:2021nmk, Freed:2022qnc}. This predicts new phases and gapless phase transitions for 1+1d systems 
\cite{Bhardwaj:2024wlr, Chatterjee:2024ych} but also in 3+1d theories  \cite{Dumitrescu:2023hbe, Antinucci:2024ltv} with non-invertible symmetries.

\section{Conclusion}
The cross-section through Formal Theory presented in this paper provides substantial evidence that the field is very much alive and kicking. Several ideas -- AdS/CFT holography, scattering amplitude methodology, compactifications of string theory -- have been central to the field for decades. 
Here the recent progress is in the details, and in the precision of the results: the computation of string amplitudes in AdS spacetimes using QFT insights, the sharpening of swampland conjectures using a refined understanding of the landscape of string compactifications, and the enormous wealth of techniques that keep getting uncovered to push the boundaries of scattering amplitudes. Modifications of some of these ideas, such as flat space holography, open up new avenues, that have seen enormous activity in the past years. 

In turn, many new ideas have been breaking through and tearing up some of our basic theoretical assumptions -- e.g.  that a symmetry need not be a group, but can be a category, and that such structures occur in condensed matter systems, but also in the Standard Model of Particle Physics. The full extent of the implications that these symmetries have is yet to be uncovered and surely will hold many future surprises.

\subsubsection*{Acknowledgements.}
I am grateful to many colleagues for discussions on this broad set of topics, and providing feedback. In particular I thank Fabrizio Caola,  Adam Kmek, Mark Mezei, Romain Ruzziconi, Akshay Yelleshpur Srikant for discussions on some of the subjects further afield to my main research area, and Miguel Montero and Tobias Hansen for inspiring talks at Strings 2024. Thanks to Romain, Adam and Miguel for comments on the draft. If there are errors these are all mine. I thank Fabio Apruzzi, Ibou Bah, Lakshya Bhardwaj, Federico Bonetti, Lea Bottini, Dan Pajer, Apoorv Tiwari, Alison Warman and Jingxiang Wu, for collaborations on many exciting papers on Generalized Symmetries, that were mentioned here. 
This work is supported by the UKRI Frontier Research Grant, underwriting the ERC Advanced Grant ``Generalized Symmetries in Quantum Field Theory and Quantum Gravity” and in part by  STFC grant ST/X000761/1.

\bibliography{ICHEP.bib}
\bibliographystyle{JHEP}



\end{document}